\documentclass[twocolumn,aps,pra,superscriptaddress,showpacs]{revtex4}
\usepackage{epsfig}
\begin{document}
\title{Reversibility of continuous-variable quantum cloning}
\author{Radim Filip} 
\affiliation{Department of Optics, Palack\'{y} University, 
17. listopadu 50, 77200 Olomouc, Czech Republic}

\author{Jarom\'{\i}r Fiur\'{a}\v{s}ek}
\affiliation{Quantum Information and Communication, 
Ecole Polytechnique, CP 165, \\
Universit\'{e} Libre de Bruxelles, 1050 Bruxelles, Belgium }
\affiliation{Department of Optics, Palack\'{y} University, 
17. listopadu 50, 77200 Olomouc, Czech Republic}

\author{Petr Marek}
\affiliation{Department of Optics, Palack\'{y} University, 
17. listopadu 50, 77200 Olomouc, Czech Republic}

\date{\today}
\begin{abstract} 
We analyze a reversibility of optimal Gaussian $1\rightarrow 2$ quantum cloning 
of a coherent state using only local operations on the clones and 
classical communication between them and propose a feasible 
experimental test of this feature. Performing Bell-type homodyne measurement on one 
clone and anti-clone, an arbitrary unknown input 
state (not only a coherent state) can be restored 
in the other clone by applying
appropriate local unitary displacement operation. We generalize this concept to a partial LOCC reversal 
of the cloning and we show that this procedure converts the symmetric cloner 
to an asymmetric cloner. Further, we discuss a distributed LOCC 
reversal in optimal $1\rightarrow M$ Gaussian cloning of coherent states 
which transforms it to optimal  $1\rightarrow M'$ cloning for $M'<M$.
Assuming the quantum cloning as a possible eavesdropping 
attack on quantum communication link, the reversibility can be 
utilized to improve the security of the link even after the attack.  
\end{abstract}
\maketitle

\section{Introduction}

The linearity of quantum mechanics imposes a fundamental constraint on the
processing of quantum information that has no classical counterpart. Perhaps the
most famous example is the no-cloning theorem which states that an unknown
quantum state cannot be copied  \cite{nocloning}. However, although exact 
copying is forbidden, one can still perform an approximate cloning.
A quantum cloning machine is a device that 
with the help of an ancilla produces two or more approximate copies of an
unknown quantum state. The problem of designing the optimal cloning 
transformations that maximize the fidelity of the clones 
has attracted a considerable amount of attention during recent 
years and  optimal universal cloning machines for qubits \cite{UQCM} 
and qudits \cite{Buzek98} have been found. 
Experimentally,  quantum cloning of polarization state of a photon was 
accomplished by means of stimulated parametric down-conversion
\cite{Lamas-Linares02,DeMartini02} and also by projecting two photonic qubits 
on a symmetric subspace with the help of a balanced beam splitter \cite{Ricci03}.
Recently, the concept of cloning has been 
also extended to the domain of continuous variables where Gaussian cloning 
machine for coherent states has been proposed 
\cite{Cerf00,Braunstein01,Fiurasek01}.  
 
The cloning operation can be represented by a unitary transformation on the 
total system consisting of the original state, blank copies and ancillas. 
This operation can be trivially reversed by applying  inverse
unitary transformation to the total system. In contrast, in this paper, 
we analyze a different {\em LOCC reversibility} \cite{Gregoratti03} 
of cloning which uses only 
\emph{local operations} on the clones and classical communication between them.   
The LOCC reversibility of universal $1\rightarrow 2$ cloning of qubits 
was theoretically discussed by Bruss \emph{et al.} \cite{Bruss01} who
shown that the reversal can be accomplished by  local operations on the 
two clones and classical communication by two different  methods: 
(a) probabilistically, using separate measurements on ancilla and clone 
with unit fidelity but only with maximal success rate $1/3$ or (b) 
deterministically, using complete Bell-state analysis on one clone and ancilla 
followed by an appropriate unitary operation on the other clone, similarly as 
in the quantum teleportation \cite{Bennett93}.

In this paper, we extend the concept of LOCC reversibility 
to continuous-variable (CV) quantum cloning of coherent states where a feasible
experimental test of {\em deterministic} LOCC reversibility of the cloning 
can be envisaged. Note that a different kind of deterministic LOCC 
reversibility was recently  analyzed in the context of CV quantum non-demolition 
measurements in Ref. \cite{QNDreverse}. Here, we shall demonstrate 
that the  Braunstein-Kimble teleportation scheme \cite{BKtele,BKteleexp}
realizes deterministic LOCC reversal of the optimal Gaussian symmetric and even
asymmetric $1\rightarrow 2$ quantum cloning of coherent states \cite{Braunstein01,
Fiurasek01}. The method is general and works for an arbitrary initial state 
(not only for coherent states). We point out the differences between  the LOCC 
reversibility of the  qubit and CV quantum cloning.  Then we will show that it 
is possible to perform only a \emph{partial reversal} of the $1\rightarrow 2$ 
cloning by means of LOCC operations. Finally, we extend the
LOCC reversibility  to the $1 \rightarrow M$ cloning.

Besides being of fundamental interest, the full and partial reversal of the
cloning via LOCC operations may also find practical applications in quantum
communication with continuous variables. For instance, the asymmetric cloning
of coherent states represents the optimal individual Gaussian eavesdropping 
attack on the quantum key distribution protocol with coherent states, 
homodyne detection and direct reconciliation \cite{Grosshans02} and
it also saturates the information exclusion principle 
when used as an eavesdropping attack on quantum key distribution protocol 
based on squeezed states proposed in Ref. \cite{inf}.
Imagine that the eavesdropper Eve keeps one clone and the anti-clone and sends
the other clone to Bob, see Fig. 1. This attack may significantly reduce the rate of the secret
key generation between the two 
communicating parties Alice and Bob and, eventually, it may prevent Alice and 
Bob from generating any secret key at all. However, if Eve subsequently 
agrees to cooperate with Bob (perhaps for some fee) she can supply to
Bob the data that will allow him to correct the outcomes of his homodyne  measurements 
such that Alice and Bob will be able to distill secret key from the corrected
data. As we shall see, the data transmitted during the reversal procedure
does not carry any information about the signal, so in
principle Eve does not learn anything about the secret key by performing the
reversal operation.

\begin{figure}[t]
\centerline{\psfig{width=0.95\linewidth,angle=0,file=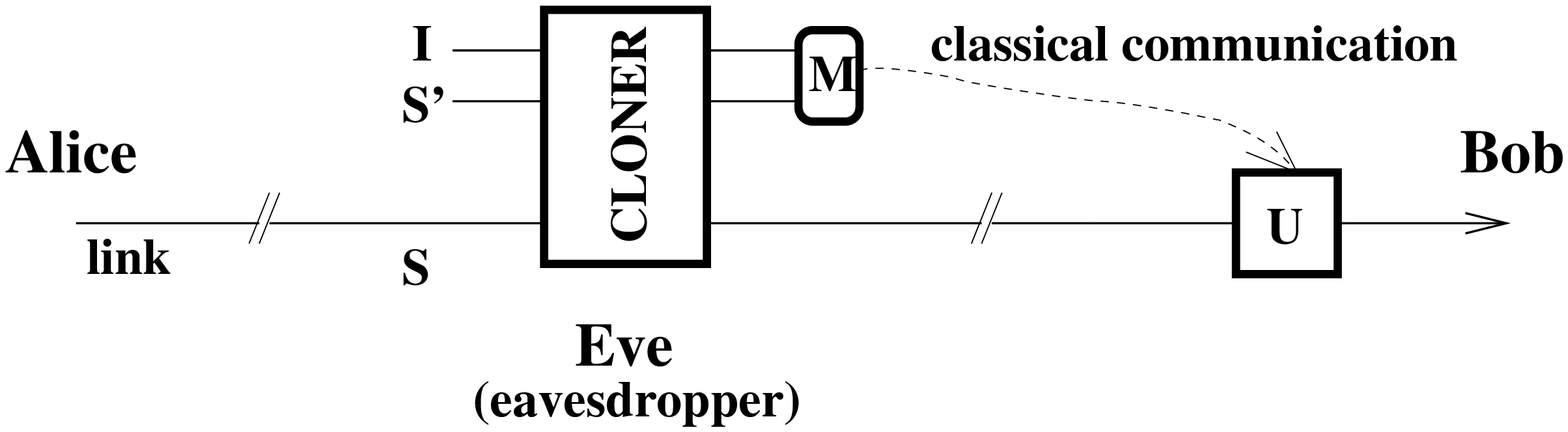}}
\caption{Schematic of the LOCC reversal of the cloning. After the attack via a cloning
machine, the eavesdropper Eve performs a  suitable measurement $M$
on one clone $S'$ and anti-clone $I$. Bob can enhance the fidelity of his clone 
and consequently  the security of the link using only a classical 
information from Eve.}
\end{figure}

The paper is organized as follows. In Section II we discuss the full LOCC reversal of the 
asymmetric $1\rightarrow 2$ cloning of coherent states. In Section III we extend 
this procedure to only a partial LOCC reversal that effectively converts 
symmetric cloning machine to the asymmetric one.  A generalized
reversal procedure for  $1\rightarrow M$ cloning machines is devised
in Section IV. Finally, the conclusions are drawn in Section V.

\section{Total reversal of asymmetric cloning}

The proposed setup for the complete LOCC reversal of asymmetric cloning of coherent states
is schematically depicted in Fig. 2. The setup can be divided into several
blocks including the cloning, the Bell measurement on one clone and the
anti-clone, displacement of the state of the other clone and, finally, the 
analysis of the state of this clone by means of homodyne detection in order 
to confirm the success of the reversal. In what follows, we discuss each 
of these operations in detail.

The optimal and universal (in the sense that all coherent states are cloned
with the same fidelity)  Gaussian $1\rightarrow 2$ asymmetric quantum cloning 
of coherent states is based on  phase-insensitive amplification of the 
input single-mode optical field \cite{Fiurasek01}. The phase-insensitive 
amplification of coherent states was demonstrated with a good fidelity in the 
back-action evading measurements \cite{QND,2xQND}. For this 
reason, we propose an implementation of the asymmetric cloning using 
this available experimental technique. 

A scheme of the cloning device is situated in the upper left part of Fig. 2. 
First we split a fraction of the input coherent state 
in linearly polarized mode $S$ to the orthogonal polarization mode $S'$  
with the use of a half-wave plate WP1 aligned 
at angle $\theta_{1}$ with respect to the optical axis. 
The polarization mode $S'$ is subsequently deflected to different spatial mode
by a polarizing beam splitter PBS1. 
The mode $S$ is then amplified with controllable amplitude gain $G$ 
in a type-II phase-matched nonlinear crystal (KTP). The signal $S$ and idler 
$I$ modes are two orthogonally polarized beams co-propagating through the crystal.  
The pump P for the crystal is prepared by frequency doubler $2\omega$ 
from the master laser  source  L and it is injected and extracted by two 
dichroic mirrors DM1 and DM2. The intensity of pumping controls 
the gain $G$  of the amplification. 
The idler beam containing the anti-clone (phase-conjugated state) 
after amplification is subtracted by polarizing beam splitter PBS2 
and replaced by  the mode $S'$ having the same polarization.    
The two orthogonally polarized modes $S$, $S'$ are combined on half-wave
plate WP2 aligned at angle $\theta_{2}$ and finally $S$ and $S'$ are 
separated to different spatial  modes at polarizing beam splitter PBS3. 
The optimal Gaussian asymmetric cloning machine for coherent states is obtained
by  setting the angles $\theta_{1},\theta_{2}$ and the gain $G$ as follows
\cite{Fiurasek01}:
\begin{eqnarray}
\tan\theta_{1}&=& \sqrt{2}\sinh(\gamma), \nonumber \\
G&=&\sqrt{2}\cosh(\gamma), \\
\tan\theta_{2}&=& -\exp(2\gamma), \nonumber 
\end{eqnarray}
where $\gamma$ characterizes the asymmetry of the cloner.

\begin{figure}[b]
\centerline{\psfig{width=0.99\linewidth,angle=0,file=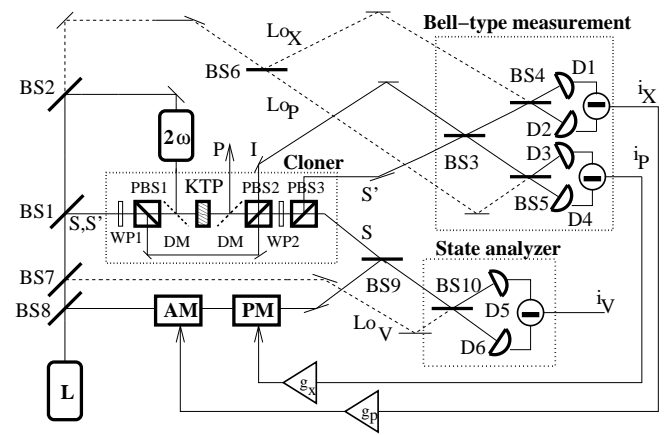}}
\caption{Setup for test of deterministic LOCC reversibility of CV quantum cloning: 
L -- laser, BS1-BS10 -- beam splitters, 
$2\omega$ -- frequency doubler, PBS1-3 -- polarization beam splitters, WP1,2 -- half-wavelength plates, 
DM -- dichroic mirrors,  
KTP -- type-II collinear phase-matched KTP nonlinear crystal, D1-D6 -- detectors, 
AM,PM -- electro-optical modulators, g$_x$,g$_p$ -- electronic amplifiers.}
\end{figure}

It is most convenient to perform the calculations in the Heisenberg picture,
where the optimal Gaussian $1\rightarrow 2$ asymmetric cloning of coherent 
states is described by the following linear canonical transformation of 
the annihilation operators of the three involved modes \cite{Fiurasek01}:
\begin{eqnarray}\label{cloner}
  A'_{S}  &=& A_{S} + \frac{\exp(-\gamma)}{\sqrt{2}}\left(A_{S'} + A_{I}^{\dag}\right), \nonumber\\
  A'_{S'} &=& A_{S} - \frac{\exp(\gamma)}{\sqrt{2}}\left(A_{S'} - A_{I}^{\dag} \right), \nonumber\\
  A'_{I}  &=& \sqrt{2}\cosh\gamma A_{I}-\sqrt{2}\sinh\gamma A^{\dag}_{S'} +
  A^{\dag}_{S}.
\end{eqnarray}
The annihilation operators $A_{i}=(X_{i}+iP_{i})/\sqrt{2}$, $i=S,S',I$  
satisfy the commutation relations $[A_i,A_i^{\dag}]=1$. The fidelity $F_{j}$ between
the input coherent state and an output state of the $j$th clone can be 
expressed as
\begin{equation}
\label{Fidelityj}
F_{j}=\frac{1}{1+\bar{n}_{\mathrm{ch},j}},
\end{equation}
where $\bar{n}_{\mathrm{ch},j}$ is the mean number of chaotic photons 
in $j$th clone. Specifically, the asymmetric cloning produces two replicas 
with the fidelities \cite{Fiurasek01}
\begin{equation}
F_{S}=\frac{2}{\exp(-2\gamma)+2}, \qquad F_{S'}=\frac{2}{\exp(2\gamma)+2}.
\label{Fidelitiesasymmetric}
\end{equation} 
The optimal symmetric Gaussian cloner with $\gamma=0$ \cite{Cerf00,Braunstein01,Fiurasek01}
yields two equivalent clones  with fidelity $2/3$ with respect to the initial 
coherent state. In case of symmetric cloning $\theta_1=0$ and
the proposed setup can be simplified because we can remove WP1 and also PBS1.
Interestingly, the asymmetric cloning machine can be obtained from the symmetric
cloner by feeding the input ports $S'$ and $I$ with two-mode squeezed 
vacuum \cite{Grosshansthesis}.   

We  now propose an experimentally feasible total reversal of the asymmetric 
cloning.  Suppose that Eve possesses one clone in mode $S'$ 
and also the idler mode $I$ (i.e. the anti-clone) while Bob has the other 
clone in mode $S$, see Fig. 1. They would like to restore in Bob's mode $S$ the original state sent 
by Alice  only by  means of LOCC operations. It follows from Eq. (\ref{cloner}) 
that Bob must remove the noise in the  operator $A'_S$ which is represented 
by the term   $Y=e^{-\gamma}(A_{S'}+A^{\dag}_{I})/\sqrt{2}$. 
Eve can measure the non-Hermitian operator $Y$ by Bell-type detection on 
modes $S'$ and $I$ similarly as in the CV quantum teleportation \cite{BKtele}. 
In this way Eve can simultaneously measure two commuting quadratures 
$X'_{-}=(X_{S'}^\prime-X_{I}^\prime)/\sqrt{2}$ and
$P'_{+}=(P_{S'}^\prime+P_{I}^\prime)/\sqrt{2}$. It is easy to check that 
$Y=-X_{-}^\prime-iP_{+}^\prime$.

The Bell measurement is carried out by an eight-port homodyne 
detector which consists of a balanced beam splitter BS3 and 
two balanced homodyne detectors (BHDs), as is depicted in Fig.~2. 
The homodyne detectors involve balanced beam splitters BS4, BS5 and
four detectors D1-D4. The necessary local oscillators $Lo_{X},Lo_{P}$  
are derived from the common master laser beam. 
After the measurements, the operators $X_{-}^\prime$ and $P_{+}^\prime$ 
``collapse'' to random measured  values $x_{-}^\prime$ and $p_{+}^\prime$ 
hence the measured value of $Y$ is $y=-x_{-}^\prime-ip_{+}^\prime$
which is physically encoded in a pair  of photo-currents $i_{X},i_{P}$ 
leaving the homodyne detectors. These photo-currents represent 
the classical message that has to be transmitted from Eve to Bob 
to reverse the cloning attack. It is clear from Eq. (\ref{cloner}) that 
the noise arising in the signal mode $S$  can be fully eliminated by the 
coherent displacement on mode $S$
\begin{equation}\label{disp}
A''_{S} =  A'_{S} - y,  
\end{equation}
which finishes the reversal. In the experiment, a suitable displacement can be accomplished by overlapping 
the mode $S$ with a coherent state modulated in amplitude and phase 
by electro-optical modulators AM and PM on a strongly unbalanced 
(e.g. $99:1$) beam splitter BS9, as illustrated in Fig. 2.
To achieve the required displacement, the photo-currents are pre-amplified 
in $g_x$ and $g_p$ in a proper way. Note that the reversal operation is 
essentially the quantum teleportation of continuous-variable state 
\cite{BKtele,BKteleexp}. 
Finally, a tomography of reversed state can be performed in the state analyzer. 

Since $A_{S}^{\prime \prime}=A_{S}$ the  LOCC  reversibility of 
CV asymmetric cloning
can be performed deterministically for {\em arbitrary} input
state. Moreover, if a thermal state with a high temperature 
is injected  in the auxiliary inputs we are able to precisely 
restore any input state of mode $S$ although both clones exhibit only  
a negligible fidelity and all the information about the input state is
stored in the intermodal correlations. 
In addition, this reversal can be performed even 
\emph{a-posteriori}, after the  homodyne measurement on  Bob's clone, simply
by data manipulation corresponding to the displacement (\ref{disp}). 
This is important for the quantum cryptographic applications, as  discussed in
the Introduction.

The reversibility of quantum cloning for qubit \cite{Bruss01} 
and CV coherent states seems to be analogical. However, there are important 
differences between the reversibility of qubit and CV quantum cloning 
arising from  a different distribution of the discrete and continuous 
variable quantum information. Mainly, it was shown in \cite{Bruss01} 
that we  can conditionally restore the 
initial state with unit fidelity (albeit with non-unit success rate) 
only by separate  local measurements on one clone and ancilla,
which is impossible for the CV cloning. 
Further, it was shown in Ref. \cite{Bruss01} 
that sufficient information to conditionally restore the input state in the
anti-clone can be gained  by measurement performed on the two clones. 
The same is clearly not true in the present case because 
in order to recover the original state of the signal mode $S$ in the idler mode
$I$ we would have to measure the ``noise'' operator 
$Z=\sqrt{2}(A_I\cosh\gamma -A_{S'}^\dagger\sinh\gamma )+A_S^\dagger-A_S$. 
However, the Hermitian and anti-Hermitian
parts of $Z$ do not commute hence it is impossible to measure $Z$.

\section{Partial reversal of asymmetric cloning}

So far, we have considered only full reversal of the cloning transformation 
that perfectly restores the original state on the receiver's side. 
In this section, we extend this procedure to a partial optimal reversal. 
We shall show that by means of LOCC operations it is possible to convert 
the output of the asymmetric cloning machine (\ref{cloner}) 
to the output of another optimal Gaussian asymmetric
cloning machine with a different asymmetry parameter $\gamma'$.

\begin{figure}
\psfig{figure=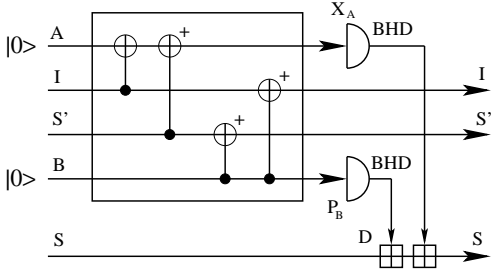,width=0.75\linewidth}
\caption{Logical quantum network schematic of the LOCC partial reversal of 
$1\rightarrow 2$ cloning of coherent states. The two clones are contained in 
modes $S$ and $S'$, respectively, $I$ is the anti-clone and $A$ and $B$ 
denote auxiliary modes. 
The lines connecting the modes indicate CV QND operations with control 
mode indicated by full circle and target mode by empty circle, respectively.
The quadratures $X_A$ and $P_{B}$ are measured by two balanced 
homodyne detectors (BHD) and the  quadratures of the Bob's mode $S$ are 
displaced accordingly. See text for details.}
\end{figure}

The main idea behind the partial reversal of the cloning is to perform a
a partial quantum  non-demolition (QND) measurement of the two commuting 
noise operators  $X'_{-}$ and $P'_{+}$. This allows us to reduce to some extent 
the noise in Bob's clone while the states of modes $S'$ and $I$ are also partially
preserved although some noise is added due to the back-action of the QND
measurement. The whole procedure is schematically depicted in terms of
the quantum network in Fig.~3. Two auxiliary modes 
initially prepared in the vacuum state and labeled $A$ and $B$  
are used on Eve's side.
The information on the measured quadratures is transferred onto the 
quadratures $X_{A}$ and $P_{B}$, respectively, via a sequence of four QND
interactions
\begin{equation}\label{QNDcoupling}
\begin{array}{lcl}
X'_{j}=X_j, & & X'_{k}=X_{k}+\kappa X_{j},  \\
P'_{j}=P_{j}-\kappa P_{k}, &\qquad &P'_{k}=P_{k},
\end{array}
\end{equation}
where $j$ and $k$ are labels of the two interacting modes. 
We say that $j$ is the control mode (denoted by full dot in
Fig.~3) and $k$ is the target mode (denoted by open circle with 
cross in Fig.~3).
The inverse of (\ref{QNDcoupling}) is obtained by change of the sign of 
the coupling constant,  $\kappa\rightarrow -\kappa$, which is indicated 
in Fig.~3 by a dagger. 
The value of $\kappa$ determines the strength of the measurement of the noise
quadratures, i.e. the asymmetry of the two clones after the partial reversal.
Recently, the QND interaction has been realized between two polarization
modes of optical field \cite{QND,2xQND} and also between polarization 
of bright beam and collective-spin degree of freedom of a macroscopic atomic 
ensemble \cite{Kuzmich00,Julsgaard01,Schori02}.

Since the QND interaction is phase-sensitive it is most convenient to work 
with quadrature operators. After the sequence of the four QND interactions 
depicted in Fig.~3, we have
\begin{eqnarray}
X''_{S'}&=& X_{S} - \frac{\exp(\gamma)}{\sqrt{2}}(X_{S'} - X_{I}) -\kappa X_{B}, \nonumber \\
P''_{S'}&=&P_{S}-\frac{\exp(\gamma)}{\sqrt{2}}(P_{S'}+	P_{I})+\kappa P_{A}, \nonumber \\
X''_{I}&=&\sqrt{2}\cosh(\gamma) X_{I}
-\sqrt{2}\sinh(\gamma) X_{S'} + X_{S}-\kappa X_{B},   \nonumber\\
P''_{I}&=& \sqrt{2}\cosh(\gamma) P_{I}
+\sqrt{2}\sinh(\gamma) P_{S'}-P_{S}-\kappa P_{A},  \nonumber\\
X''_{A}&=&X_{A}+\frac{\kappa\exp(-\gamma)}{\sqrt{2}} (X_{S'}+X_{I}), \nonumber \\
P''_{B}&=&P_{B}+\frac{\kappa\exp(-\gamma)}{\sqrt{2}} (P_{I}-P_{S'}).
\end{eqnarray}
The operators $X''_{A}$ and $P''_{B}$ are measured by means of two balanced homodyne detectors and
the measurement results $x_{A}$ and $p_{B}$ are communicated to Bob who displaces 
the quadratures $X'_{S}$ and $P'_{S}$ of his clone as follows:
\begin{equation}\label{partialdisplacement}
X''_{S} =X'_{S}-g x_{A}, \qquad P''_{S}= P'_{S}+g p_{B}.
\end{equation}
The ``teleportation gain'' $g$ should be chosen such as to minimize the fluctuations of the 
output quadratures $X''_{S}$ and $P''_{S}$. 
In the case of optimal Gaussian asymmetric cloning of coherent states 
we find that the optimal gain reads
\begin{equation}\label{gain}
g=\frac{\kappa}{\exp(2\gamma)+\kappa^2}
\end{equation}
and we  effectively get 
\begin{eqnarray}
X''_{S} &=& X_{S} +
\frac{e^{\gamma}}{e^{2\gamma}+\kappa^2}\frac{1}{\sqrt{2}}(X_{S'} + X_{I}) -
\frac{\kappa}{e^{2\gamma}+\kappa^2} X_{A}, 
\nonumber \\
\qquad P''_{S} &=& P_{S} + \frac{e^{\gamma}}{e^{2\gamma}+\kappa^2}\frac{1}{\sqrt{2}}(P_{S'} - P_{I}) 
+\frac{\kappa}{e^{2\gamma}+\kappa^2} P_{B}.
\nonumber 
\end{eqnarray}

We claim that with the gain (\ref{gain}) the partial reversal procedure 
produces two optimal Gaussian asymmetric clones in modes $S$ and $S'$.
To prove this we calculate the fidelities of the clones in modes $S$ and $S'$
after partial reversal, assuming that the mode $S$ was initially in coherent
state while $S'$ and $I$ were in vacuum state. With the help 
of formula (\ref{Fidelityj}) we obtain after a simple algebra
\begin{eqnarray}
F_{S}&=&\frac{2[\kappa^2+\exp(2\gamma)]}{2[\kappa^2+\exp(2\gamma)]+1},\nonumber\\
F_{S'}&=&\frac{2}{2+\exp(2\gamma)+\kappa^2}.
\end{eqnarray}
These are exactly the fidelities (\ref{Fidelitiesasymmetric}) 
with an effective asymmetry parameter $\gamma'$ given by
\begin{equation}\label{gamma'}
\gamma'=\frac{1}{2}\ln[\exp(2\gamma)+\kappa^2].
\end{equation}
This confirms that our procedure yields the same output as the
optimal Gaussian asymmetric cloner (\ref{cloner}) with $\gamma=\gamma'$.

\begin{figure}
\psfig{figure=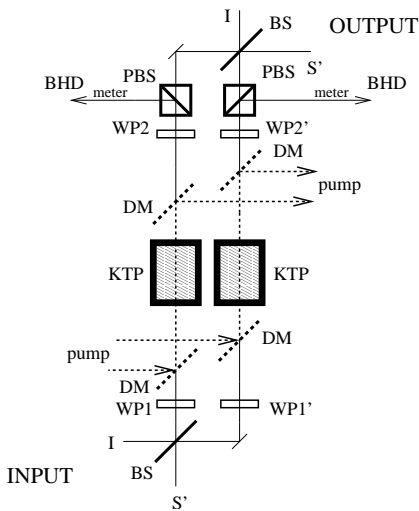,width=0.65\linewidth}
\caption{Partial CV Bell-state measurement. KTP is a type-II phase-matched 
nonlinear crystal,  BS are balanced beam splitters, 
DM is a dichroic mirror, WP1, WP1', WP2, WP2' are half-wave plates aligned 
at evasion angle, PBS is polarizing beam  splitter, 
BHD are balanced homodyne detectors. See text for more details. }
\end{figure}

The QND coupling required for the implementation of the partial reversal 
of the cloning of coherent
states naturally emerges as an effective interaction describing the off-resonant coupling of
light with the collective spin degree of freedom of an ensemble of atoms. So, if the clones and
anti-clones were stored in the atomic memories \cite{Schori02}, one could perform the 
required QND measurements of $X'_{-}$ and $P'_{+}$ simply by sending two beams 
of light through the two ensembles, followed by homodyne detection on the output
beams, similarly as in the experiment where two ensembles were entangled by 
interaction with a light pulse that passed through them \cite{Julsgaard01}.

An all-optical implementation of the partial reversal is a bit more involved
but nevertheless feasible. The QND measurement of the optical field 
quadrature can be  implemented  using single  type-II phase-matched
nonlinear crystal  sandwiched between two half-wave plates aligned at evasion angle, 
and the signal and the meter are two orthogonally polarized beams 
co-propagating through the device \cite{QND}.  
For a particular gain-dependent orientation of the 
half-wave plates, this device accomplishes the QND interaction (\ref{QNDcoupling}). 
The setup depicted in Fig.~3 can be simplified to a scheme that 
requires only two QND interactions 
and two balanced beam splitters. The resulting all-optical setup is depicted 
in Fig. 4. The modes $S'$ and $I$ are first combined on a
balanced beam splitter. 
Two QND devices are used for partial measurement of the $X$ quadrature of the
first output mode and the $P$ quadrature of the second mode.
After the QND interactions, the modes $S'$ and $I$ are re-combined on second balanced beam 
splitter while the quadratures $X_{A}$ and $P_{B}$ are measured by two homodyne detectors. 
Since the experiments with two QND measurements in the setup 
have been already reported \cite{2xQND}, this experimental 
implementation of the partial Bell type measurement for 
the continuous variables appears to be feasible.

\section{Distributed reversal of 1$\rightarrow$(M,M-1) CV cloning}

Let us now consider an extension of the LOCC reversal
procedure to the symmetric $1\rightarrow (M,M-1)$ cloning machine that
produces $M$ clones and $M-1$ anticlones from a single replica 
\cite{Braunstein01,Fiurasek01,Cerf00',Cerf01}. 
We assume that $M-1$ clones and $M-1$ anti-clones are distributed among $M-1$
eavesdroppers such that each eavesdropper possesses one clone and one anticlone.
The remaining single clone is sent to Bob, 
as is depicted in Fig.~5.  After the distribution, 
a part $L \leq M-1$ of eavesdroppers  decide to collaborate with Bob 
who would like to increase the fidelity of his clone. 
We show that it is sufficient if each collaborating eavesdropper performs 
local Bell type measurement on her clone and anti-clone and sends the
measurement results to Bob via  classical communication channel.
As a result, Bob can obtain a clone with fidelity 
corresponding to optimal Gaussian $1\rightarrow M-L$ 
cloning only using LOCC operations.  

The symmetric $1\rightarrow (M,M-1)$ cloner is based on the 
phase-insensitive amplification, similarly as the 
$1\rightarrow 2$ cloner considered in Sec. II. 
After the amplification  with gain $G=\sqrt{M}$, the signal mode 
is split into $M$ spatially separated modes by $M \times M$ balanced coupler 
that can be described by  the following transformation of annihilation 
operators 
\begin{equation}
A'_{S,j}=\frac{A'_S}{\sqrt{M}}+\sum_{k=1}^{M-1} u_{jk}A_k,
\end{equation}
where $j=1,\ldots, M$ and $A'_S$ is annihilation operator of signal mode 
leaving the amplifier which is sent to the $M$th input port of the coupler.
The modes in the remaining $M-1$ input ports characterized by
annihilation operators  $A_k$ are assumed to be in vacuum state.
We define $u_{jM}=1/\sqrt{M}$ and the matrix $u_{jk}$ is unitary. 
Similarly, the idler mode is also split into $M-1$ separate modes by the 
same technique. The symmetric $1\rightarrow (M,M-1)$ cloner 
can be thus described by the following transformation, 
\begin{eqnarray}\label{1-(M,M')}
A'_{S_j}&=& A_{S}+\sqrt{\frac{M-1}{M}}A^{\dag}_{I}+\sum_{k=1}^{M-1}u_{jk} A_{k},
\nonumber\\
A'_{I_l}&=& A_{S}^{\dag}+\sqrt{\frac{M}{M-1}}A_{I}+\sum_{k=2}^{M-1} v_{lk} B_k,
\end{eqnarray}
where $l=1,..., M-1$, $B_k$ are annihilation operators of auxiliary modes 
which are assumed to be in vacuum states. The matrix $v_{lk}$ is
unitary if we define $v_{l1}=1/\sqrt{M-1}$.
Each clone in the $M$ 
signal modes $A_{S_j}^\prime$ produced by this cloner exhibits the same fidelity 
\begin{equation}
F_{1\rightarrow M}=\frac{M}{2M-1}.
\end{equation}

\begin{figure}
\centerline{\psfig{width=8.0cm,angle=0,file=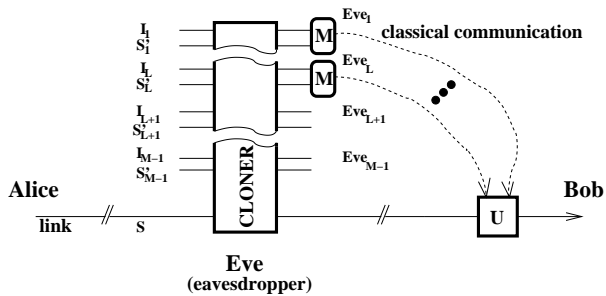}}
\caption{Schematic of the LOCC reversal of the cloning for multiple 
eavesdroppers.  After joint attack, eavesdroppers Eve$_1,\ldots,$Eve$_L$
perform separate measurements $M$ on their pairs of clones $S'_{i}$ and 
anti-clones $I_{i}$. Bob can enhance the fidelity of his clone 
and, consequently, the  security of the link using only 
a classical information received from Eve$_{1},\ldots,$Eve$_{L}$ .}
\end{figure}

Let $1\leq L\leq M-1$ be the number of eavesdroppers willing to collaborate with Bob. 
The partial reversal of the $1\rightarrow (M,M-1)$ cloning can be achieved by 
$L$-fold eight-port homodyne measurement separately performed on pairs of 
modes $S_j$, $I_j$, where $j=1,\ldots,L$, see Fig. 5. Similarly as before,
it is a measurement of two commuting
two-mode quadratures on every pair, which can be also conveniently 
interpreted as  measuring the non-hermitian operator 
$Y_{j}=A'_{S'_{j}}-A^{\prime\dagger}_{I_{j}}$ yielding a set of 
complex numbers $y_{j}=x_{j}+ip_{j}$. If all these measurement results 
are available to Bob, he can displace his mode $S_M$
\begin{equation}\label{displacement}
A''_{S_M}=A'_{S_M}+g\sum_{j=1}^{L} y_{j},
\end{equation}
where $g$ is electronic gain adjusted properly so as to maximally enhance 
the fidelity of the Bob's clone. A simple algebra reveals that 
\begin{eqnarray}
Y_{j}=-\frac{A_{I}^{\dagger}}{\sqrt{M(M-1)}}+\sum_{k=1}^{M-1}u_{jk}A_{k}
-\sum_{k=2}^{M-1}v_{jk}^{\ast}B^{\dagger}_{k}.
\end{eqnarray}
The mean number of chaotic photons in the clone $S_M$ after the displacement 
(\ref{displacement})
can be easily calculated if we take into account that all modes $A_k$ and $B_k$
are initially in vacuum or coherent state. One obtains 
\begin{equation}
\bar{n}_{\mathrm{ch}}=\frac{M-1}{M}-\frac{2gL}{M}+\frac{g^2L^2}{M(M-1)}
+ g^{2}\sum_{j,j'=1}^{L}\sum_{k=2}^{M-1} v_{jk}^{\ast}v_{j'k}.
\label{nbarM}
\end{equation}
The triple summation in Eq. (\ref{nbarM}) can be evaluated by noting that 
$v_{jk}$ is unitary and $v_{j1}=1/\sqrt{M-1}$. The sum over $k$ reads
$\sum_{k=2}^{M-1}v_{jk}^{\ast}v_{j'k}=\delta_{jj'}-1/(M-1)$. On inserting this
back into Eq. (\ref{nbarM}) and summing over $j$ and $j'$ we finally obtain
\[
\bar{n}_{\mathrm{ch}}=\frac{M-1}{M}-\frac{2gL}{M}+\frac{g^2L^2}{M(M-1)}
+ g^{2}\left(L-\frac{L^2}{M-1}\right).
\]
The maximum fidelity is achieved by minimizing the noise 
$\bar{n}_{\mathrm{ch}}$ and the  optimal gain is given by
\[
g_{\mathrm{opt}}= \frac{1}{M-L}.
\]
For this gain  we get $\bar{n}_{\mathrm{ch}}=(M-L-1)/(M-L)$. 
The corresponding fidelity of coherent-state clone after the reversal
\begin{equation}\label{FM}
F''_{S}=\frac{M-L}{2(M-L)-1},
\end{equation}
is exactly the fidelity of optimal symmetric Gaussian $1\rightarrow M-L$
cloning of coherent states. A complete restoration of the original occurs 
only for $L=M-1$, i.e. when all eavesdroppers collaborate with Bob.

\section{Conclusions}

In summary, we have analyzed the LOCC reversibility of quantum cloning 
of coherent states and proposed  a feasible setup for an experimental test. 
We have proved  that the Braunstein-Kimble teleportation scheme for continuous variables 
\cite{BKtele} can be used for deterministic LOCC reversal of the 
asymmetric $1\rightarrow 2$ cloning machine \cite{Braunstein01,Fiurasek01} 
for arbitrary input state.  
Furthermore, it is possible to perform only 
a partial LOCC reversal based on a partial Bell type measurement which effectively 
converts the symmetric cloner to asymmetric one. This principle 
can be extended also to reversal of $1\rightarrow (M,M-1)$ cloning machines. 

The prospects for the experimental demonstration of 
the total LOCC reversal of the symmetric $1\rightarrow 2$ 
cloning seem to be very good. First, a high-quality amplification 
of the coherent signal was experimentally demonstrated \cite{QND}. 
Second, in the CV teleportation experiment \cite{BKteleexp}, 
the achieved fidelity ($F\approx 0.64$) 
is close to the maximal value corresponding to the finite amount of
entanglement shared between Alice and Bob. These spectacular experimental 
achievements strongly suggest that with the present day technology it should 
be possible to  demonstrate the increase of cloning fidelity from $2/3$ to 
the value near unity via the LOCC reversal procedure.

Quantum cloning machines represent frequently considered
individual coherent eavesdropping attacks on quantum key distribution
protocols \cite{Grosshans02,inf,attack}.  
An eavesdropper Eve uses two imperfect clones of the state sent by Alice.
Eve transmits one clone to Bob, and measures the other when the measurement 
basis is revealed during the reconciliation procedure. 
Due to the reversibility of quantum cloning, 
it is possible in principle to recover a full security of the quantum 
communication link only  by local operations on clones and classical 
communication between Eve and Bob in the case that  Eve has decided 
to collaborate with Bob in a time after her attack on the link 
and before the measurement on her clone and anti-clone.

\begin{acknowledgments}
We would like to thank Nicolas Cerf and Ladislav Mi\v{s}ta, Jr. for stimulating and
helpful discussions. The work was supported by the project LN00A015 
and CEZ: J14/98  of the Ministry of Education of Czech Republic, by the 
EU grant under QIPC project IST-1999-13071 (QUICOV) and project 202/03/D239 of 
The Grant Agency of the Czech Republic. JF also acknowledges support
from the EU under project CHIC (IST-2001-32150). 
\end{acknowledgments}


\end{document}